%% file: main_Allerton23_v1.tex
\documentclass[twocolumn,conference]{IEEEtran}
\usepackage{latexsym}
\usepackage{bm}
\usepackage[dvips]{graphics}
\usepackage{graphicx,subfigure}
\usepackage{psfrag}
\usepackage{amsfonts,amsmath,amssymb,hyperref,amsthm}
\usepackage{algorithm,algorithmic}
\usepackage{cite}

\DeclareMathAlphabet{\mathbb}{U}{bbold}{m}{n}

\usepackage{dsfont,color,subfigure}

\input{defns.tex}

\newtheorem{question}{Question}

\newtheorem{theorem}{Theorem}

\newtheorem{corollary}{Corollary}

\IEEEoverridecommandlockouts
\begin{document}

\title{Theoretical Analysis of Binary Masks in Snapshot Compressive Imaging Systems}
\author{Mengyu Zhao\thanks{MZ is with the Electrical and computer Engineering Department of Rutgers University. E-mail: mz524@soe.rutgers.edu} \and 
Shirin Jalali \thanks{SJ is with the Electrical and computer Engineering Department of Rutgers University. E-mail: shirin.jalali@nokia-bell-labs.com}} 
\maketitle
 
\newcommand{\p}{\mathds{P}}
\newcommand{\mb}{\mathbf{m}}   
\newcommand{\bb}{\mathbf{b}}

\begin{abstract}
Snapshot compressive imaging (SCI) systems have gained significant attention in recent years. While previous theoretical studies have primarily focused on the performance analysis of Gaussian masks, practical SCI systems often employ binary-valued masks. Furthermore, recent research has demonstrated that optimized binary masks can significantly enhance system performance. In this paper, we present a comprehensive theoretical characterization of binary masks and their impact on SCI system performance. Initially, we investigate the scenario where the masks are binary and independently identically distributed (iid), revealing a noteworthy finding that aligns with prior numerical results. Specifically, we show that the optimal probability of non-zero elements in the masks is smaller than 0.5. This result provides valuable insights into the design and optimization of binary masks for SCI systems, facilitating further advancements in the field. Additionally, we extend our analysis to characterize the performance of SCI systems where the mask entries are not independent but are generated based on a stationary first-order Markov process. Overall, our theoretical framework offers a comprehensive understanding of the performance implications associated with binary masks in SCI systems.

\end{abstract}

\maketitle


\section{Introduction}

\begin{figure}[h]
\begin{centering}
\includegraphics[width=7.5cm]{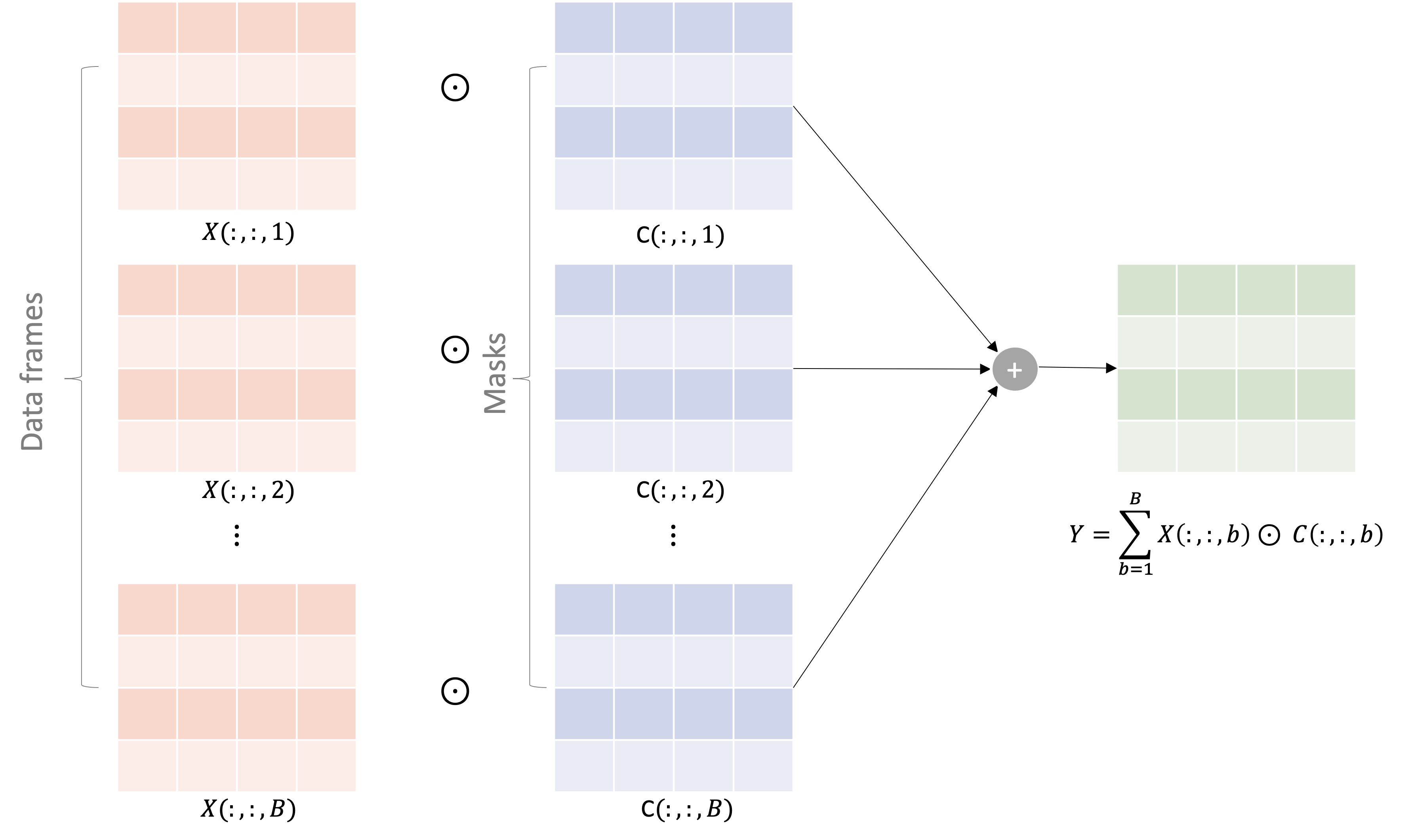}
\par\end{centering}
\caption{SCI encoding function: For $b=1,\ldots,B$, frame $b$ and mask $b$ are represented  by  $X(:,:,b)$ and $C(:,:,b)$, respectively. The single  2D measurement frame is generated as $\sum_{b=1}^BX(:,:,b)\odot C(:,:,b)$.}\label{fig:sys-model}
\end{figure}

Snapshot compressive imaging (SCI) refers to  imaging systems that are designed to map a high-dimensional (HD) 3D data cube into a 2D image through hardware. (Refer to Fig.~\ref{fig:sys-model} for a schematic model of SCI systems encoding function.) The desired HD 3D data cube is then recovered from the 2D projection using  proper algorithms. The  motivation behind developing  SCI solutions is to make the data acquisition phase more efficient. For instance, a key application of SCI is in hyperspectral imaging (HSI). HSI is an emerging technology with a wide range of applications,  from medicine to astronomy, see e.g. \cite{liu2011tongue,lu2014medical,hege2004hyperspectral,kurz2013close,peyghambari2021hyperspectral}. The key challenge with classic HSI solutions is that they rely on scanning the image either in space or along the wavelengths. This makes the HSI process slow and costly. To address this challenge, a snapshot compressive hyperspectral imaging solution has been proposed that can dramatically speed up the process by capturing all the information in a single snapshot \cite{gehm2007single}. 

In recent years, numerous hardware solutions for SCI have been proposed for various applications (for an overview, refer to  \cite{yuan2021snapshot}).  As shown in Fig.~\ref{fig:sys-model}, the encoding operation of SCI systems can be modeled as highly under-determined linear inverse problems with specialized sensing matrices. Various methods have been proposed in the literature for solving such inverse problems., e.g. see \cite{saideni:hal-03593023,qiao2020deep,meng2020gap,9495194,wang2022spatialtemporal}. 
While SCI systems are under-determined linear inverse problems, the specialized structure of their sensing matrices on one hand, and the complex structure of the input data on the other hand, prevents results from compressed sensing to be directly applicable to such systems. Therefore, for theoretical analysis of such systems new tools and techniques are required. Such a theoretical analysis is performed in \cite{jalali2019snapshot}, for the case where the corresponding  linear mapping can be modeled as a highly sparse matrix with its non-zero entries distributed as i.i.d.~Gaussian. The analysis of  \cite{jalali2019snapshot} theoretically shows that recovery of the signal from SCI measurements is indeed feasible. However, practical SCI systems often employ binary-valued (or finite-valued) masks. Also, in many practical cases, the masks corresponding to different frames are not independent and are instead (sometimes randomly \cite{llull2013coded}) shifted versions of each other. These raise the following questions.
\begin{question}
For binary-valued masks, can we theoretically characterize the performance of the SCI system in terms of the statistical properties of the masks, e.g., the probability of non-zero entries, or the correlation between adjacent (in-frame or out-of-frame) values? If the answer is positive, can we use these theoretical results to theoretically  optimize the performance of the system?
\end{question}
\begin{question}
How does the dependency between the masks used for different frames affect the achievable SCI performance? For a specified type of dependence between the masks, e.g., randomly shifted masks, can we optimize the initial mask, such that the achievable performance is optimized?
\end{question}
The goal of this paper is to address these questions. To achieve this goal we adopt  the compression-based optimization, compressible signal pursuit (CSP), initially proposed in \cite{jalali2016compression} and later utilized in \cite{jalali2019snapshot} for the theoretical analysis of SCI systems. Within this framework, we make the following main contributions:
\begin{enumerate}
\item Theoretical characterization of CSP optimization performance for SCI recovery under i.i.d. binary masks. Our analysis reveals that the probability of non-zero entries minimizing the achieved distortion is less than 0.5.
\item Theoretical characterization of CSP optimization performance for scenarios where non-zero entries in each frame exhibit dependence, following a binary first-order Markov process. In this case, we assume that the non-zero entries of different masks are independent.
\item Investigation of the impact of dependency across frames by studying cases where entries across frames are dependent, aiming to model the effect of mask dependence.

\end{enumerate}



%

\subsection{Related work}

Recent research has focused on optimizing masks to enhance the performance of SCI systems. Trained sensing binary masks have been explored, demonstrating notable  improvements over random mask designs \cite{iliadis2020deepbinarymask}. These optimized binary masks have a nonzero element probability of around 40 percent and exhibit smooth variations. Deep unfolding networks have also been employed to simultaneously reconstruct hyperspectral images and optimize mask designs, resulting in preserved image structure and optimal sampling \cite{Zhang_2022_CVPR}. Furthermore, a comparison between random masks and optimized masks, validated with hardware prototypes, supports the benefits of optimization \cite{koller2015high}. Other approaches include incorporating apertures for mask multiplexing \cite{Zhang_2021} and using end-to-end networks to jointly optimize masks and networks, leading to improved loss function and peak signal-to-noise ratio (PSNR) performance \cite{9105237}. Collectively, these studies highlight the impact of mask optimization in enhancing SCI system performance.

\subsection{Notations}
Vectors are denoted by bold letters, such as $\xv$ and $\yv$. For a matrix $\Xv\in\mathds{R}^{n_1\times n_2}$, ${\rm Vec}(\Xv)$ denotes the  vector in $\mathds{R}^n$, $n=n_1\times n_2$, formed by concatenating the columns of $\Xv$. $\odot$ denotes the Hadamard matrix product operator defined as follows. For $\Av,\Bv\in\mathds{R}^{n_1\times n_2}$, $\Yv=\Av\odot \Bv$ is defined such that $Y_{ij}=A_{ij}B_{ij}$, for all $i,j$. Sets are denoted by calligraphic letters, such as $\Ac,\Bc$. For a finite set $\Ac$, $|\Ac|$ denotes the size of $\Ac$

\subsection{Outline}
The mathematical models of SCI systems encoding and decoding operations are described in Section \ref{sec:SCI}. Section \ref{sec:CSP} reviews the idea of compression-based methods for solving SCI inverse problems. The main theoretical results of the paper are presented in Section \ref{sec:main} and the proofs are presented in Section \ref{sec:proofs}.  Section \ref{sec:conclusion} concludes the paper. 
\section{Problem statement}\label{sec:SCI}
The goal of an SCI system is to recover a 3D data cube from its 2D projection, while knowing the mapping. More precisely,  let $\Xv \in \mathbb{R}^{n_1 \times n_2 \times B}$ denote the desired 3D data cube. An SCI system maps $\xv$ to a single measurement frame  $\Yv \in \mathbb{R}^{n_1\times n_2}$. In many SCI systems, such as HS SCI \cite{gehm2007single} and video SCI \cite{llull2013coded},  the mapping from $\Xv$ to $\Yv$ can be modeled as a linear system such that \cite{llull2013coded,Wagadarikar09CASSI}, $\Yv = \sum_{b=1}^B \Cv_b\odot \Xv_b + \Zv$, where $\Cv\in \mathbb{R}^{n_1 \times n_2 \times B}$  and $\Zv \in \mathbb{R}^{n_1 \times n_2 }$ denote the sensing kernel  (mask) and the additive noise, respectively. Here, $\Cv_b = \Cv(:,:,b)$ and $\Xv_b = \Xv(:,:,b) \in \mathbb{R}^{n_1 \times n_2}$ represent the $b$-th sensing kernel (mask) and the corresponding signal frame, respectively; Here, $\odot$ denotes the Hadamard or element-wise product.

To simplify the mathematical representation of the system, we vectorize each frame as  $\xv_b={\rm Vec}(\Xv_b)\in\mathbb{R}^n$ with $n = n_1 n_2$.  Then, we vectorize the data cube  $\Xv$ by concatenating the $B$ vectorized frames into a column vector $\xv \in \mathbb{R}^{n B}$ as
\begin{equation} \label{Eq:xv1toB} 
\xv = \left[\begin{array}{c}
\xv_1\\
\vdots\\
\xv_B
\end{array}\right].
\end{equation}
Similarly, we define  $\yv = \text{Vec}(\Yv) \in \mathbb{R}^{n}$ and $\zv= \text{Vec}(\Zv) \in \mathbb{R}^{n}$. Using these definitions, the measurement process defined in Fig.~\ref{fig:sys-model} can also be expressed  as 
\begin{align}
\yv = \Hv \xv + \zv.\label{eq:SCI-model}
\end{align}
The sensing matrix $\Hv\in\mathbb{R}^{n\times nB}$, is a highly sparse matrix that is formed by the  concatenation of $B$  diagonal matrices as
\begin{equation}\label{Eq:Hmat_strucutre}
\Hv = [\Dv_1,...,\Dv_B],
\end{equation}
where, for  $b =1,\dots B$, $\Dv_b = \text{diag}(\text{Vec}(\Cv_b)) \in {\mathbb R}^{n \times n}$. 
The goal of a SCI recovery algorithm is to recover the data cube $\xv$ from undersampled measurements $\yv$, while having access to the sensing matrix (or mask) $\Hv$.

\section{Compression-based SCI recovery}\label{sec:CSP}
One of the key challenges in theoretical analysis of SCI systems is developing a mathematical model for the structure of 3D data cubes, such as videos or HS images. On approach to address this issue is to use the idea of compression-based recovery, which was initially proposed in \cite{jalali2016compression} in the context of compressed sensing.  In that case, it can be shown that at least in cases where the minimum achievable sample rate  is known, compression-based methods are able to achieve it \cite{RezagahJ:17-IT}. Inspired by this idea, in \cite{jalali2019snapshot}, the first theoretical analysis of SCI systems was performed using data compression codes for capturing the source structure. 

Compression codes designed for a class of signals are designed to take advantage of the structure of the signals in that class to represent it as efficiently as possible. The key idea of using compression codes for solving inverse problems is to use the compression code as a black-box that implicitly takes advantage of signal structure.  In the following we briefly review some key definitions related to compression codes defined for a given class of HD data cubes. 

  Consider a compact set $\Qc\subset\mathbb{R}^{n  B}$.  Each signal $\xv\in\Qc$,
consists of $B$ vectors (frames) $\{\xv_1\ldots\xv_B\}$ in $\mathbb{R}^n$.  A lossy compression code of rate $r$ for $\Qc$  is characterized by its encoding mapping $f$, where $f: \Qc \rightarrow \{1,2,\ldots, 2^{Br}\},$ and $g: \{1,2\ldots2^{Br}\}\to \mathbb{R}^{nB}$.
For $\xv\in\Qc$,  $\tilde{\xv} = g(f(\xv))$ denotes the reconstruction corresponding to $\xv$. The {\emph {  distortion}} between  $\xv$ and its reconstruction  $\hat{\xv}$ is defined as 
\begin{equation}
 d(\xv,\hat{\xv})\triangleq   \|\xv-\hat{\xv}\|_2^2.\label{eq:def-delta-B}
\end{equation}
The compression code $(f,g)$ is characterized by its rate $r$ and distortion  $\delta$ defined as 
\[
\delta=\sup_{\xv\in\Qc}  d(\xv,g(f(\xv))).
\]
Moreover, the defined encoder and decoder pair, $(f,g)$, correspond to a codebook  ${\cal C}$ defined as
\begin{equation}
\Cc=\{g(f(\xv)):\; \xv\in\Qc\}.  \label{eq:codebook}
\end{equation}
Note that $|\Cc|\leq 2^{Br}$.

Consider the problems of SCI defined in Section \ref{sec:CSP}. To recover $\xv$ from underdetermined measurements $\yv$, we need to take advantage of the structure of $\xv$. However, as explained earlier, the desired mathematical model of the structure needs to capture both intra- on inter-frame dependencies, which makes designing such models inherently very complex. One approach to address this issue and provide a theoretical analysis is SCI systems is to adopt the idea of compression-based recovery. The key advantage of this approach is that instead of explicitly expressing the structure, it will be captured through a compression  code, and the performance is determined by the key parameters of the compression code, i.e., its rate $r$ and distortion $\delta$. 

Given a class of signal denoted by a $\Qc\subset\mathbb{R}^{nB}$, and a rate-$r$ distortion-$\delta$  compression code $(f,g)$, the compressible signal pursuit (CSP) optimization recovers $\xv\in\Qc$ from measurements $\yv\in\mathbb{R}^n$ defined in \eqref{eq:SCI-model}, as follows
\begin{align}
\hat{\xv}=\arg\min_{\cv\in \Cc}\|\yv-\sum_{i=1}^B\Dv_i\cv_i\|_2^2.\label{eq:CSP}
\end{align}
The performance of \eqref{eq:CSP} is theoretically characterized in \cite{jalali2019snapshot} for the case where the diagonal  entries of $\Dv_1,\ldots,\Dv_B$ are i.i.d.~Gaussian. In this paper, inspired by used  in practical SCI systems, we focus on the case of binary-valued masks, and under various distributions characterize the performance of \eqref{eq:CSP}. 

\section{Characterization of effect of masks}\label{sec:main}

In this section, we present our main theoretical results on the performance of SCI systems under different settings of binary masks. Our goal is to address the questions we raised before on how the statistical properties and dependencies of binary masks impact the performance of SCI systems, and whether it is possible to optimize the masks to achieve better performance. We discuss our findings in three distinct settings, each corresponding to different mask characteristics and their effects on the system's performance.

\subsection{i.i.d.~Bernoulli masks}
As the first scenario, we focus on the masks entries are i.i.d.~ and binary-valued. There are two key questions we want to address in this setting: Is recovery still feasible? If so, what is the optimal value of $p$,  $p=\P(D_{ij}=1)$,  that minimizes the achieved distortion between the signal and its SCI reconstruction?
\begin{theorem}\label{thm:1} 
Consider $\Qc\subset\mathbb{R}^{nB}$, where   for all $\xv\in\Qc$, $\|\xv\|_{\infty}\leq\frac{\rho}{2}$. Let $\Cc$ denote the codebook corresponding to a rate-$r$ distortion-$\delta$ lossy compression code for signals in $\Qc$.  Assume that  $\Dv_1\ldots\Dv_B$  are such that $\Dv_i={\rm diag}(D_{i1}\ldots D_{in})$, $i=1,\ldots,B$, where the diagonal entires of the  matrices drawn independently  i.i.d.~ ${\rm Bern}(p)$. For $\xv\in\Qc$ and $\yv=\sum_{i=1}^B\Dv_i\xv_i$ let $\hat{\xv}$ denote the solution of \eqref{eq:CSP}. Choose free parameter $\epsilon>0$. Then,
\begin{equation}
\frac{1}{nB}\|\xv-\hat{\xv}\|_2^2\leq(1+\frac{Bp}{1-p})({\delta\over nB})+{\rho^2\epsilon\over (p-p^2)},\label{eq:main-thm1-result}
\end{equation}
with a probability larger than $1-2^{Br+1}\exp(-\frac{n\epsilon^2}{2B^2})$. Moreover, for fixed parameters $(n,B,\epsilon,\rho)$, the bound in \eqref{eq:main-thm1-result} is minimized at some $p^*$, where $p^*<{1\over 2}$. 
\end{theorem}

Note that as $p\to 0$ or $p\to 1$, the bound in \eqref{eq:main-thm1-result} grows without bound. This is consistent with the fact that we cannot  expect recovery from  all-zero or  all-one masks. On the other hand, for $p=0.5$, Theorem \ref{thm:1} guarantees that 
\[
\frac{1}{nB}\|\xv-\hat{\xv}\|_2^2\leq(1+B)({\delta\over nB})+{4\rho^2\epsilon},
\]
with probability larger than $1-2^{Br+1}\exp(-\frac{n\epsilon^2}{2B^2})$. Moreover, it states that the optimal $p^*$ is smaller than $0.5$, which means that the optimal bound is tighter than this result. This is consistent with the results from the literature, e.g.~ \cite{iliadis2020deepbinarymask}, that show through various types of algorithmic optimizations that in the learned optimized binary masks  $\P(D_{ij}=1)$ is strictly smaller than $0.5$. 

One distinctive property of the studied masks compared to i.i.d.~Gaussian masks studied in the prior art is that $D_{i,j}\geq 0$, w.p. $1$. To further highlight this difference and show its potential impact on optimizing the masks, in the following corollary of Theorem \ref{thm:1}, we consider the case where instead of binary-valued, the masks take values in $\{-1,+1\}$. In that case, we see that unlike the case of binary masks, the optimal bound on the distortion  is achieved for the case where $p=0.5$ and $\E[D_{i,j}]=0$.

\begin{corollary}\label{cor:1} Consider the same setup as in Theorem 1, where instead of binary masks,  $D_{ij}\in\{-1,+1\}$ and $\{\{D_{ij}\}_{j=1}^n\}_{i=1}^B$ are i.i.d.~ such that $\P(D_{ij}=1)=1-\P(D_{ij}=-1)=p$. Then
\[
\frac{1}{nB}\|\xv-\hat{\xv}\|_2^2\leq{4(p-p^2)(1-B)+B \over 4(p-p^2)}({\delta\over nB})+{\rho^2\epsilon\over 4(p-p^2)},
\]
with a probability larger than $1-2^{Br+1}\exp(-\frac{n\epsilon^2}{2B^2})$.
Moreover, the upper bound is minimized at $p^*={1\over 2}$, which leads to $\frac{1}{nB}\|\xv-\hat{\xv}\|_2^2\leq{1 \over nB}\delta+{\rho^2\epsilon}$.
\end{corollary}

%
%
%

\subsection{Binary Markov masks: in-frame dependence}

As the first model of masks with dependent components, we consider a setting where masks corresponding to different frames are independent, but the entries of each mask are dependent and follow a first-order Markov process. More precisely, we assume that $\Dv_1,\ldots,\Dv_B$ are independent. For  $i=1,\ldots,B$, the diagonal entries of $\Dv_i$ are generated according to a stationary Markov process such that, for $j=2,\ldots,n$, 
\[
p_{D_{ij}|D_{i,1:(j-1)}}(\cdot|\cdot)=p_{D_{ij}|D_{i,j-1}}(\cdot|\cdot).
\]
Moreover, for any $i=1,\ldots,B$, and $j=2,\ldots,n$, we define the transition kernel of the (asymmetric) Markov chain as follows
\begin{align}
    \P(D_{ij}=1|D_{i(j-1)}=0)&=q_0,\nonumber\\
    \P(D_{ij}=0|D_{i(j-1)}=1)&=q_1.\label{eq:in-frame-dist}
\end{align}

%
%

%
%
%
%
%
%
%
%
%
%
%
%
%
%
%

%

\begin{figure}[h]
 \begin{center}
\includegraphics[width=7cm]{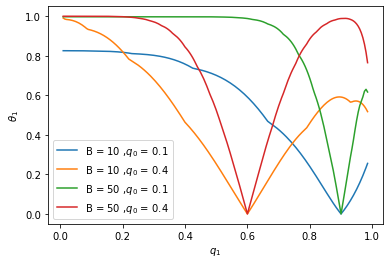}\caption{}\label{fig:2}
\end{center}
\end{figure}

To characterize the performance under the described Markov model for the masks, we use the concentration of measure results developed in \cite{kontorovich2008concentration}. For using that result, we define the contraction coefficient corresponding to the defined Markov process as 
\begin{align}
    \theta_1 
    &= \sup_{\dv',\dv''\in\Sc^B}\| p(\cdot|\dv')-p(\cdot|\dv'')\|_{\rm TV}\nonumber\\
    &= \| p_i(\cdot|\dv'={\bf 0}_B)-p_i(\cdot|\dv''={\bf 1}_B)\|_{\rm TV}\nonumber\\
    &= \frac{1}{2}[| q_0^B-(1-q_1)^B| \nonumber\\
       &\;\;\;\;\;\;\;  + \binom{B}{1}|(1-q_0)q_0^{B-1}-q_1(1-q_1)^{B-1}|\nonumber\\ 
       &\;\;\;\;\;\;\; + \binom{B}{2}|(1-q_0)^2q_0^{B-2}-q_1^2(1-q_1)^{B-2}| \nonumber\\ 
       &\;\;\;\;\;\;\;  + \cdots+ | (1-q_0)^B-q_1^B|].\label{eq:def-theta1}
\end{align}
 In \eqref{eq:def-theta1}, for $\dv,\dv'\in\Sc^B$, $p(\dv'|\dv)=\prod_{i=1}^Bp(d'_i|d_i)$ denotes the transition kernel of the defined Markov process. Fig.~\ref{fig:2} shows the value of $\theta_1$ as a function of $q_1$, for a couple of different  values of $q_0$ and $B$. 

\begin{theorem}\label{thm:2}
 Assume  that  $\Dv_1\ldots\Dv_B$  are such that $\Dv_i={\rm diag}(D_{i1}\ldots D_{in})$, $i=1,\ldots,B$, where $D_{ij}\in\{0,1\}$. Assume that $(D_{i1},\ldots,D_{in})$, $i=1,\ldots,B$, are independently generated as stationary first order Markov processes with transition probabilities described in \eqref{eq:in-frame-dist}. 
For $\xv\in\Qc$ and $\yv=\sum_{i=1}^B\Dv_i\xv_i$ let $\hat{\xv}$ denote the solution of \eqref{eq:CSP}. 
Then
\begin{equation}
\frac{1}{nB}\|\xv-\hat{\xv}\|_2^2\leq(1+\frac{Bp}{1-p})({\delta\over nB})+{\rho^2\epsilon\over p(1-p)},\label{eq:main-thm2-bd}
\end{equation}
with a probability larger than 
\[
1-(2^{Br}+1)\exp(-\frac{n\epsilon^2}{32}(1-\theta_1)^2),
\]
 where $\theta_1$ is defined in \eqref{eq:def-theta1}.
\end{theorem}

Comparing the bound in \eqref{eq:main-thm2-bd} and the one in \eqref{eq:main-thm1-result} shows that they are indeed equivalent. Therefore, similar to Theorem \ref{thm:1}, the bound is minimized for some $p^*={q_0^*\over q_0^*+q_1^*}<0.5$. On the other hand, to minimize $\exp(-\frac{n\epsilon^2}{32}(1-\theta_1)^2)$ which controls how many frames can be decoupled from each other, we need to minimize $\theta_1$ defined in \eqref{eq:def-theta1}. But we can set $\theta_1=0$, by setting $q_0^*=p^*$ and $q_1^*=1-p^*$.  It is straightforward to see that setting the parameters  $q_0$ and $q_1$ as such corresponds to making the Markov process an independent process. This is intuitively not surprising as using this setting the convergence speed of the random variables is maximized.


\subsection{Binary Markov masks: Out-of-frame dependence}
Next we consider the case where the entries of each mask are generated independently, but the mask entries corresponding to element $i$ of each frame are dependent. This is closely related to real masks used in practice where each mask can be a shifted version of the previous mask. Mathematically, we assume that $D_{1j},\ldots,D_{Bj}$ are generated according to a stationary first order Markov process such that for any $j=1,\ldots,n$ and $i=2,\ldots,B$,
\[
p_{D_{ij}|D_{1:(i-1),j}}(\cdot|\cdot)=p_{D_{ij}|D_{i-1,j}}(\cdot|\cdot),
\]
and 
\begin{align}
\P(D_{ij}=1|D_{i-1,j}=0)&=q_0\nonumber\\
\P(D_{ij}=0|D_{i-1,j}=1)&=q_1.\label{eq:out-frame-dist}
\end{align}
Assume that $q_0,q_1\leq 0.5$ and let 
\[
\alpha=1-q_0-q_1.
\]
Note that since $q_0,q_1\leq 0.5$, $\alpha\geq 0$. 
Define $B\times B$ matrix $\Lambda$ as follows
\begin{align}
\Lambda =
\begin{bmatrix}
1 & \alpha & \cdots & \alpha^{B-1} \\
    \alpha & 1& \cdots & \alpha^{B-2} \\
    \vdots & \vdots & \ddots & \vdots \\
    \alpha^{B-1} & \alpha^{B-2} & \cdots &1.
\end{bmatrix}.\label{eq:def-Lambda}
\end{align}
Let $\lambda_{\max}(\Lambda)$ and $\lambda_{\min}(\Lambda)$ denote the maximum and minimum eigenvalues of matrix $\Lambda$, respectively. 

\begin{theorem}\label{thm:3}
 Assume  that  $\Dv_1\ldots\Dv_B$  are such that $\Dv_i={\rm diag}(D_{i1}\ldots D_{in})$, $i=1,\ldots,B$, where $D_{ij}\in\{0,1\}$. Assume that $(D_{1j},\ldots,D_{Bj})$, $j=1,\ldots,n$, are independently generated as stationary first order Markov processes according to \eqref{eq:out-frame-dist}. For $\xv\in\Qc$ and $\yv=\sum_{i=1}^B\Dv_i\xv_i$ let $\hat{\xv}$ denote the solution of \eqref{eq:CSP}. 
Then, if $ \lambda_{\min}(\Lambda)>0$, 
\begin{align*}
\frac{1}{nB}\|\xv-\hat{\xv}\|_2^2\leq &{\lambda_{\max}(\Lambda)(1-p) +pB \over \lambda_{\min}(\Lambda)(1-p)}({\delta\over nB})\nonumber\\
&+{\rho^2\epsilon\over \lambda_{\min}(\Lambda) p(1-p)},
\end{align*}
 with a probability larger than $1-2^{Br+1}\exp(-\frac{n\epsilon^2}{2B^2})$.
\end{theorem}
Note that in the case where all the entries of the sensing matrix are independent, i.e., the case where $q_0+q_1=1$ and $\alpha=0$,  $ \lambda_{\max}(\Lambda)= \lambda_{\min}(\Lambda)=1$. Therefore, in that case the upper bound in Theorem \ref{thm:3} simplifies to the result of Theorem \ref{thm:1}. 

We can use Gershgorin circle theorem \cite{gershgorin1931uber} to derive upper and lower bounds on $\lambda_{\max}(\Lambda)$ and $\lambda_{\min}(\Lambda)$, respectively, and find the following corollary. 
\begin{corollary}\label{cor:thm-3}
Consider the same setup as in Theorem \ref{thm:3}. Then, for $\alpha<{1\over 3}$, 
\begin{align*}
\frac{1}{nB}\|\xv-\hat{\xv}\|_2^2\leq &{(1+\alpha)(1-p) +pB \over (1-3\alpha)(1-p)}({\delta\over nB})\nonumber\\
&+{\rho^2(1-\epsilon) \over(1-3\alpha) p(1-p)},
\end{align*}
 with a probability larger than $1-2^{Br+1}\exp(-\frac{n\epsilon^2}{2B^2})$.
\end{corollary}


\section{Proofs}\label{sec:proofs}

\subsection{Proof of Theorem \ref{thm:1}}

Let $\Tilde{\xv}=g(f(\xv))$. By assumption, the code operates at distortion $\delta$. Hence, $\| \xv-\Tilde{\xv}\|_2^2\leq\delta$.  On the other hand, since $\hat{\xv}=\arg\min_{\cv\in\Cc}\|\ \yv-\sum\nolimits_{i=1}^B\Dv_i\cv_i\|_2^2$, and $\Tilde{\xv}\in \Cc$, $\|\yv-\sum_{i=1}^B\Dv_i\hat{\xv}_i\|_2\leq\|\yv-\sum_{i=1}^B\Dv_i\Tilde{\xv}_i\|_2$. But  $\yv=\sum\nolimits_{i=1}^B\Dv_i\xv_i$. Therefore, 
\begin{equation}
\|\sum_{i=1}^B\Dv_i(\xv_i-\hat{\xv}_i)\|_2\leq\|\sum_{i=1}^B\Dv_i(\xv_i-\Tilde{\xv}_i)\|_2.\label{eq:1}
\end{equation}
Note that, for a fixed $\cv\in \Cc$,
\begin{equation}
\|\sum_{i=1}^B\Dv_i(\xv_i-\hat{\xv}_i)\|_2^2=\sum_{j=1}^n(\sum_{i=1}^BD_{ij}(x_{ij}-c_{ij}))^2
\end{equation}

Given a fixed $\xv$ and $\cv$, for $j=1, \ldots, n$, let $U_j=(\sum\nolimits_{i=1}^BD_{ij}(x_{ij}-c_{ij}))^2$.   $U_1,\ldots,U_n$ are independent  random variables and 
\begin{align}
\allowdisplaybreaks
&\E[U_j]
=\E[\sum\limits_{i=1}^B\sum\limits_{i'=1}^BD_{ij}D_{i'j}(x_{ij}-c_{ij})(x_{i'j}-c_{i'j})]\nonumber\\
&=\sum\limits_{i=1}^B\sum\limits_{i'=1 i'\neq i}^Bp^2(x_{ij}-c_{ij})(x_{i'j}-c_{i'j})+\sum\limits_{i=1}^Bp(x_{ij}-c_{ij})^2\nonumber\\
&=p^2(\sum\limits_{i=1}^B(x_{ij}-c_{ij}))^2+(p-p^2)\sum\limits_{i=1}^B(x_{ij}-c_{ij})^2. \label{eq:exp}
\end{align}

Given $\epsilon_1 > 0$ and $\epsilon_2>0$, $\xv_i\in\mathbb{R}^{n}$ and $\xv\in\mathbb{R}^{Bn}$, define events $\Ec_1$ and $\Ec_2$ as
\begin{align}
\Ec_1=\{&\frac{1}{n}\|\sum_{i=1}^B\Dv_i(\xv_i-\tilde{\xv}_i)\|_2^2
\leq\frac{p^2}{n}\|\sum_{i=1}^B(\xv_i-\tilde{\xv}_i)\|_2^2\nonumber\\
&+\frac{p-p^2}{n}\|\xv-\tilde{\xv}\|_2^2+B\rho^2\epsilon_1\}\label{eq:E1}
\end{align}
and
\begin{align}
\Ec_2=\{&\frac{1}{n}\|\sum_{i=1}^B\Dv_i(\xv_i-\cv_i)\|_2^2\geq\frac{p^2}{n}\|\sum_{i=1}^B(\xv_i-\cv_i)\|_2^2\nonumber\\
&+\frac{p-p^2}{n}\|\xv-\cv\|_2^2-B\rho^2\epsilon_2:\forall\cv\in \Cc\},
\label{eq:E2}
\end{align}
respectively. Then, conditioned on $\Ec_1\cap\Ec_2$, since $\hat{\xv}\in \Cc$ and $\Tilde{\xv}\in \Cc$, it follows from \ref{eq:1} that
\begin{align}
&\frac{p-p^2}{n}\|\xv-\hat{\xv}\|_2^2\leq\frac{p-p^2}{n}\|\xv-\Tilde{\xv}\|_2^2+\frac{p^2}{n}\|\sum_{i=1}^B(\xv_i-\tilde{\xv}_i)\|_2^2\nonumber\\
&\;\;\;\;\;\;\;\;\;\;\;\;\;\;\;\;\;\;\;\;\;\;\;\;\;\;\;\;\;+B\rho^2\epsilon_1+B\rho^2\epsilon_2\nonumber\\
&\leq\frac{p+(B-1)p^2}{n}\|\xv-\Tilde{\xv}\|_2^2+B\rho^2\epsilon_1+B\rho^2\epsilon_2,
\end{align}
where the last line follows because $\|\sum_{i=1}^B(\xv_i-\tilde{\xv}_i)\|_2^2\leq B\|\xv-\tilde{\xv}\|_2^2$. 
In the rest of the proof, we focus on bounding $P(\Ec_1^c\cup\Ec_2^c)$. \\


%

Note that since by assumption the $\ell_{\infty}$-norm of all signals in $\Qc$ are upper-bounded by $\rho / 2$,  $U_i$'s are also bounded as 
\begin{align}
U_j
&\leq\sum_{i=1}^BD_{ij}^2\cdot\sum_{i=1}^B(x_{ij}-c_{ij})^2\nonumber\\
&\leq\sum_{i=1}^B1\cdot\sum_{i=1}^B(\frac{\rho}{2}+\frac{\rho}{2})^2=B^2\rho^2.\label{eq:bound-Uj}
\end{align}


Therefore, applying the Hoeffding's inequality, 
\begin{align}
&\P(\frac{1}{n}\sum_{j=1}^nU_j\geq\frac{1}{n}\E[\sum_{j=1}^nU_j]+B\rho^2\epsilon_1)\nonumber\\
&\leq\exp(-\frac{2n^2B^2\rho^4\epsilon_1^2}{n(B^2\rho^2)^2})=\exp(-\frac{2n\epsilon_1^2}{B^2}). 
\label{eq:C1}
\end{align}
Similarly,
\begin{align}
&\P(\frac{1}{n}\sum_{i=j}^nU_j\leq\frac{1}{n}\E[\sum_{j=1}^nU_j]-B\rho^2\epsilon_2)\nonumber\\
&\leq\exp(-\frac{2n^2B^2\rho^4\epsilon_2^2}{n(B^2\rho^2)^2})=\exp(-\frac{2n\epsilon_2^2}{B^2}).
\label{eq:C2}
\end{align}
Therefore, 
\begin{equation}
\P(\Ec_1^c)
\leq\exp(-\frac{2n\epsilon_1^2}{B^2})\label{eq:E1c-prob}
\end{equation}
and, by the union bound, since  $|\Cc|\leq 2^{Br}$,
\begin{equation}
\P(\Ec_2^c)
\leq 2^{Br}\exp(-\frac{2n\epsilon_2^2}{B^2}).\label{eq:E2c-prob}
\end{equation}
Again by the union bound, $P(\Ec_1\cap\Ec_2)\geq 1-P(\Ec_1^c)-P(\Ec_2^c)$. Given $0<\epsilon<\frac{16}{3}$, the desired result follows by letting $\epsilon_1=\epsilon_2=\epsilon/2$. Plug this into (7), and we have
\begin{equation}
\begin{aligned}
\frac{p-p^2}{n}\|\xv-\hat{\xv}\|_2^2&\leq\frac{p+(B-1)p^2}{n}\|\xv-\Tilde{\xv}\|_2^2+B\rho^2\epsilon_1+B\rho^2\epsilon_2\\
&\leq\frac{p+(B-1)p^2}{n}\delta+B\rho^2\epsilon.
\end{aligned}
\end{equation}
Also, from \ref{eq:E1c-prob} and \ref{eq:E2c-prob}, for $\epsilon_1=\epsilon_2=\epsilon/2$, $P(\Ec_1\cap\Ec_2)\geq 1-\exp(-\frac{2n\epsilon_1^2}{B^2})-2^{Br}\exp(-\frac{2n\epsilon_2^2}{B^2})=1-(2^{Br}+1)\exp(-\frac{n\epsilon^2}{2B^2})$.

Finally, to finish the proof, let $f(p)=\frac{(p+(B-1)p)}{(1-p)n}\delta+{B\rho^2\epsilon\over (p-p^2)}$. Note that $f(0)=f(1)=\infty$, which is consistent with our intuition  that  all-1 or an all-0 masks are not effective. Let $p^*$ denote the value of $p\in(0,1)$ that minimizes $f(p)$, note that 
\begin{align}
f'(p)&=\frac{(B-1)(1-p)+(1+Bp-p)}{(1-p)^2n}\delta-\frac{\epsilon\rho^2(1-2p)}{p^2(1-p)^2}\nonumber\\
&={1\over (1-p)^2p^2}(\frac{\delta B p^2 }{n}-(1-2p)\epsilon \rho^2).
\end{align}
Note that $f'(0)=-{\epsilon \rho^2 \over (1-p)^2p^2}<0$ and $f'({1\over 2})={1\over (1-p)^2p^2}\frac{\delta B p^2 }{n}>0$, which implies that $p^*$ where $f'(p^*)=0$ belongs to $(0,{1\over 2})$.

\subsection{Proof of Corollary \ref{cor:1}}
The proof follows similar to the proof of Theorem \ref{thm:1}. The only difference is that, here, 
\begin{align}
 &   \E[U_j]
=\E[\sum\limits_{i=1}^B\sum\limits_{i'=1}^BD_{ij}D_{i'j}(x_{ij}-c_{ij})(x_{i'j}-c_{i'j})]\nonumber\\
&=(2p-1)^2(\sum\limits_{i=1}^B(x_{ij}-c_{ij}))^2+4(p-p^2)\sum\limits_{i=1}^B(x_{ij}-c_{ij})^2.
\end{align}

\subsection{Proof of Theorem \ref{thm:2}}
Following the same steps as  the initial steps of proof of Theorem \ref{thm:1}, we have
\[
\sum_{j=1}^n(\sum_{i=1}^BD_{ij}(x_{ji}-\hat{x}_{ij}))^2\leq\sum_{j=1}^n(\sum_{i=1}^BD_{ij}(x_{ij}-\tilde{x}_{ij}))^2
\]
Define $\dv_j=[D_{1j},\ldots,D_{Bj}].$
Note that $\dv_1,\ldots,\dv_n$ is a stationary first Markov process with state space $\Sc=\{0,1\}^B$ such that 
\[
p(\dv_i|\dv_1,\ldots,\dv_{i-1}) =p(\dv_i|\dv_{i-1}) =\prod_{j=1}^Bp(d_{ij}|d_{(i-1)j}),
\]
where $p(d_{ij}|d_{(i-1)j})$ agrees with the transition probability of the Markov chain used for generating the masks. 
Given $\xv$ and $\cv$,  for $j=1, \ldots, n$, let $\varphi(\dv_j)=(\sum\nolimits_{i=1}^BD_{ij}(x_{ij}-c_{ij}))^2$.  Unlike in the proof of Theorem \ref{thm:1}, $\varphi(\dv_1),\ldots,\varphi(\dv_n)$ are no longer independent. However, the expected values of $\varphi(\dv_j)$'s are the same as those of $U_j$'s, because the dependencies are in-frame. Therefore, 
\[
\E[\varphi(\dv_j)]
=p^2(\sum\limits_{i=1}^B(x_{ij}-c_{ij}))^2+(p-p^2)\sum\limits_{i=1}^B(x_{ij}-c_{ij})^2. \nonumber
\]
Similar to the proof of Theorem \ref{thm:1}, define events $\Ec_1$ and $\Ec_2$, as \eqref{eq:E1} and \eqref{eq:E2}, respectively. 
To bound $\P((\Ec_1\cap\Ec_2)^c)$, we need  to show the concentration of $\sum_{j=1}^n\varphi(\dv_j)$ around its expected value. To achieve this goal, we use a result from \cite{kontorovich2008concentration}, which is explained in Appendix \ref{app:a}. To employ Theorem \ref{thm:app-a-2}, note that since the Markov chain is assumed to be stationary, $\theta_1=\theta_2=\ldots =\theta_{n-1}$. Therefore,
\begin{align}
    M_n &= \max_{1\leq i\leq n-1}(1+\theta_i+\theta_i\theta_{i+1}+\cdots+\theta_i\cdots\theta_{n-1})\nonumber\\
    &=1+\theta_1+\theta_1^2+\theta_1^3+\cdots+\theta_1^{n-1}\nonumber\\
    &=\frac{1-\theta_1^n}{1-\theta_1}.
\end{align}




To use  Theorem \ref{thm:app-a-2} stated in Appendix \ref{app:a}, let $c$ denote the Lipschitz coefficient  of function $\varphi:\Sc\to\mathbb{R}$, defined earlier. Then, we have
\begin{align}
&\P\big(\;\frac{1}{n}\sum_{j=1}^n\varphi(D_j)\geq\frac{1}{n}\E\left[\varphi(D_j)\right]+B\rho^2\epsilon_1\big)\nonumber\\
&\leq\exp(-\frac{n^2B^2\rho^4\epsilon_1^2}{2nc^2M_n^2})\nonumber\\
&\leq\exp(-\frac{nB^2\rho^4\epsilon_1^2}{2c^2}(1-\theta_1)^2),\label{eq:bd-E1}
\end{align}
and
\begin{align}
&\P\big(\;\frac{1}{n}\sum_{j=1}^n\varphi(D_j)\leq\frac{1}{n}\E\left[\varphi(D_j)\right]-B\rho^2\epsilon_2\big)\nonumber\\
&\leq\exp(-\frac{n^2B^2\rho^4\epsilon_2^2}{2nc^2M_n^2})\nonumber\\
&\leq\exp(-\frac{nB^2\rho^4\epsilon_2^2}{2c^2}(1-\theta_1)^2),\label{eq:bd-E2}
\end{align}
where in deriving both bounds we have used the fact that  $M_n={1-\theta_1^n \over 1-\theta_1}\leq {1\over 1-\theta_1}.$
To bound the Lipschitz constant $c$, note that for and $\dv_j,\dv'_j\in\{0,1\}^B$, we have 
\begin{align}
    &|\varphi(\dv_j)-\varphi(\dv'_j)|\nonumber\\
    &=|(\sum_{i=1}^BD_{ij}(x_{ij}-c_{ij}))^2-(\sum_{i=1}^BD_{ij}'(x_{ij}-c_{ij}))^2|\nonumber\\
    &=|\sum_{i=1}^B (D_{ij} + D_{ij}')(x_{ij}-c_{ij})| \cdot |\sum_{i=1}^B (D_{ij} - D_{ij}')(x_{ij}-c_{ij})|\nonumber\\
    &\stackrel{\rm (a)}{\leq} 2B\rho^2 d_H(\dv_j,\dv'_j),
\end{align}
where (a) follows because for all $\xv\in\Qc$, $\|\xv\|_{\infty}\leq {\rho \over 2}$. This implies that $c\geq 2B\rho^2$. 
Finally, setting $\epsilon_1=\epsilon_2=\epsilon/2$, and noting that  $| \Cc |\leq 2^{Br}$ yields the desired result. 

%
%
%



\subsection{Proof of Theorem \ref{thm:3}}
Again we follow the same steps as  the initial steps of proof of Theorem \ref{thm:1} to derive $\sum_{j=1}^n(\sum_{i=1}^BD_{ij}(x_{ji}-\hat{x}_{ij}))^2\leq\sum_{j=1}^n(\sum_{i=1}^BD_{ij}(x_{ij}-\tilde{x}_{ij}))^2.$
As in the proof of Theorem \ref{thm:2},  define $\dv_j=[D_{1j},\ldots,D_{Bj}].$ Unlike the proof of Theorem \ref{thm:2}, here $\dv_1,\ldots,\dv_n$ are independent and identically distributed. Again similar to the proof of Theorem \ref{thm:1}, given $\xv$ and $\cv$, for $j=1, \ldots, n$,  define 
\[
U_j(\xv,\cv)=(\sum\nolimits_{i=1}^BD_{ij}(x_{ij}-c_{ij}))^2.
\]  Note that   $U_1(\xv,\cv),\ldots,U_n(\xv,\cv)$ are independent  random variables. Moreover, they are positive and  bounded with the same upper bound as the one derived in \eqref{eq:bound-Uj}. Therefore, we can still apply the Hoeffding's inequality and derive \eqref{eq:C1} and \eqref{eq:C2}. The key difference now is that computing $\E[U_j(\xv,\cv)]$ is more complex as the entries of $\dv_j$ are not independent.  

To compute $\E[U_j(\xv,\cv)]$, define $\mu_{ij}=x_{ij}-c_{ij}$. Also, note that as before, $\E[D_{ij}^2]=\E[D_{ij}]=p$. Moreover,
\begin{align}
\E[U_j(\xv,\cv)]&=\E[(\sum_{i=1}^BD_{ij}(x_{ij}-c_{ij}))^2]\nonumber\\
&=\sum_{i_1=1}^B\sum_{i_2=1}^B\E[D_{i_1j}D_{i_2j}]\mu_{i_1j}\mu_{i_2j}.\label{eq:Uj-step1}
\end{align}
Without loss of generality, assume that $i_1<i_2$. Then, $\E[D_{i_1j}D_{i_2j}]=\P(D_{i_1j}=D_{i_2j}=1)=\P(D_{i_1j}=1)\P(D_{i_2j}=1|D_{i_1j}=1)$. To compute $\P(D_{i_2j}=1|D_{i_1j}=1)$, we need to compute $(i_2-i_1)$-th order transition probability of the Markov chain. 
%
The transition kernel of the Markov chain can be written as 
\[P=\begin{bmatrix}
    1-q_0 & q_0\\
    q_1 & 1-q_1\\
\end{bmatrix}.
\]
Let $Q=\begin{bmatrix}
    1 & -q_0\\
    1 & q_1\\
\end{bmatrix}$, and let $\alpha= 1-q_0-q_1.$
Then,
\begin{gather}
    \Pi=Q\begin{bmatrix}
        1 & 0\\
        0 & \alpha\\
    \end{bmatrix}Q^{-1}
\end{gather}
Using this representation, the $k$-th order transition probability of this Markov chain can be written as 
\begin{gather}
     \Pi^k=\frac{1}{q_0+q_1}
    \begin{bmatrix}
        q_1 & q_0\\
        q_1 & q_0\\
    \end{bmatrix}
    +
    \frac{\alpha^k}{q_0+q_1}
    \begin{bmatrix}
        q_0 & -q_0\\
        -q_1 & q_1\\
    \end{bmatrix}.
\end{gather}
Therefore, for $k=1,\ldots,n-i$
\[
\P(D_{(i+k)j}=1|D_{ij}=1)={q_0+\alpha^k q_1\over q_0+q_1}=p+(1-p)\alpha^k. 
\]
Thus,
\begin{align}
\E&[U_j(\xv,\cv)]=\sum_{i_1}^B\sum_{i_2}^B\E[D_{i_1j}D_{i_2j}]\mu_{i_1j}\mu_{i_2j}\nonumber\\
&=\sum_{i_1}^B\sum_{i_2}^Bp(p+(1-p)\alpha^{|i_1-i_2|})\mu_{i_1j}\mu_{i_2j}\nonumber\\
&=p^2(\sum_{i}^B\mu_{ij})^B+p(1-p)\sum_{i_1}^B\sum_{i_2}^B\alpha^{|i_1-i_2|}\mu_{i_1j}\mu_{i_2j}\nonumber\\
&=p^2(\sum_{i}^B\mu_{ij})^B+p(1-p)\muv_j^T \Lambda \muv_j,
\label{eq:E-Uj-exact}
\end{align}
where $\muv_j=[\mu_{1j},\ldots,\mu_{Bj}]^T$ and $\Lambda$ is defined in \eqref{eq:def-Lambda}. Therefore, $\E[U_j(\xv,\cv)]$ can be upper- and lower-bounded as
\begin{align}
\E[U_j(\xv,\cv)]&\leq p^2(\sum_{i=1}^B\mu_{ij})^2+p(1-p)\lambda_{\max}(\Lambda) \|\muv_j\|_2^2,\label{eq:Uj-lb}
\end{align}
and
\begin{align}
\E[U_j(\xv,\cv)]&\geq p^2(\sum_{i=1}^B\mu_{ij})^2+p(1-p)\lambda_{\min}(\Lambda) \|\muv_j\|_2^2.\label{eq:Uj-ub}
\end{align}
Define,
\[
 \Ec_1=\{\sum_{j=1}^n U_j(\xv,\cv)\geq \sum_{j=1}^n \E[U_j(\xv,\cv)]-B\rho^2\epsilon_1,\;\forall \cv\in\Cc\},
\]
and
\[
 \Ec_2=\{\sum_{j=1}^n U_j(\xv,\tilde{\xv})\leq \sum_{j=1}^n \E[U_j(\xv,\tilde{\xv})]+B\rho^2\epsilon_2,\},
\]
respectively. As explained earlier, we the bounds in  \eqref{eq:C1} and \eqref{eq:C2} still hold here too. Therefore, the lower bound on $\Ec_1\cap\Ec_2$ is the same as before. But conditioned on $\Ec_1\cap\Ec_2$, employing the bounds in \eqref{eq:Uj-lb} and \eqref{eq:Uj-ub}, it follows that  
\begin{align*}
&\frac{p(1-p)\lambda_{\min}(\Lambda)}{n}\|\xv-\hat{\xv}\|_2^2\nonumber\\
&\leq\frac{p(1-p)\lambda_{\max}(\Lambda)}{n}\|\xv-\Tilde{\xv}\|_2^2+\frac{p^2}{n}\|\sum_{i=1}^B(\xv_i-\tilde{\xv}_i)\|_2^2\nonumber\\
&\;\;\;\;+B\rho^2\epsilon_1+B\rho^2\epsilon_2\nonumber\\
&\leq\frac{p(1-p)\lambda_{\max}(\Lambda)+p^2B}{n}\delta+B\rho^2\epsilon,
\end{align*}
where the last line follows by setting $\epsilon_1=\epsilon_2={\epsilon\over 2}$. 

\subsection{Proof of Corollary \ref{cor:thm-3}}
According to the Gershgorin circle theorem \cite{gershgorin1931uber}, since all the diagonal entries of $\Lambda$ are equal to one, every eigenvalue of $\Lambda$ lies within at least one of the Gershgorin discs. These discs are all centered at one and have radii equal to 
\[
r_i=\sum_{j\neq i}\Lambda_{ij},
\]
for $i=1,\ldots,B$. Therefore,
\[
1- \max_i r_i \leq \lambda_{\min}(\Lambda)\leq \lambda_{\max}(\Lambda)\leq 1+\max r_i.
\] 
But 
\[
\max r_i\leq 2(\alpha+\alpha^2+\ldots)={2\alpha\over 1-\alpha}.
\]
Therefore,
\[
{1-3\alpha\over 1-\alpha} \leq \lambda_{\min}(\Lambda)\leq \lambda_{\max}(\Lambda)\leq {1+\alpha\over 1-\alpha}.
\]
Inserting these bounds in the result of Theorem \ref{thm:3} yields the desired result.

\section{Conclusion}\label{sec:conclusion}

In this paper, we have theoretically studied the performance of SCI systems under different types of binary masks. Prior art had characterized the theoretical performance of SCI systems under i.i.d.~Gaussian masks. However, in practice the masks are rarely i.i.d.~Gaussian. In many applications, the masks are binary-valued. Moreover, there have been results in the literature on optimizing binary masks such that  the performance of SCI system is optimized. In this paper, we have characterized the performance of SCI systems under three different models for  binary masks. Our results theoretically confirm the observations in the literature that for i.i.d.~binary masks to optimize the performance the probability of 1s should be smaller than $0.5$. 
\begin{appendices}
\section{Concentration Inequalities for Dependent Random Variables}
\label{app:a}

Here, we briefly review the key results of \cite{kontorovich2008concentration} which we  use in proving Theorem \ref{thm:2}. Consider a collection of random variables $(X_i)_{1\leq i\leq n}$ taking values in a countable space $\Sc$. Assume $X_i$ are the coordinate projections defined on the probability space $(\Sc,\Fc,\P)$. Let $\Sc^n$ be equipped with the Hamming metric $d : \Sc^n \times \Sc^n \rightarrow [0, \infty)$, defined as  $d_{\rm H}(x, y)~\dot{=}\sum_{i=1}^n \mathbb{1}_{\{x_{i}\not=y_{i}\}}.$
Let $\E$ denote expectation with respect to $\P$. Also, given two random variables $Y$ and $Z$, let $\Lc(Z | Y = y)$ denote the conditional distribution of $Z$ given $Y = y$.

For the metric probability space denoted as $(\Sc^n,d_{\rm H},\P)$, we define the following mixing coefficients. For $1\leq i<j\leq n$, let
\begin{align}
\textstyle    \Bar{\eta}_{ij}  ~\dot{=} \sup_{{y^i-1}\in\Sc^{i-1},w,\hat{w}\in\Sc} \eta_{ij}(y^{i-1},w,\hat{w}),
\end{align}
where
\begin{align}
\textstyle    &\eta_{ij}(y^{i-1},w,\hat{w})\nonumber\\
\textstyle    &\dot{=}\|\Lc(X^n_j|X^i=y^{i-1}w)-\Lc(X^n_j|X^i=y^{i-1}\hat{w})\|_{\rm TV}.
\end{align}
Note that $\eta_{ij}(y^{i-1},w,\hat{w})\leq 1$. 
Define  an  $n\times n$ upper-triangular matrix  $\Delta_n$ (it only considers the previous value in a sequence) such that 
\begin{align}
\textstyle    (\Delta_n)_{ij} = 
    \begin{cases}
    1, & \text{if}~i=j\\
    \Bar{\eta}_{ij}, & \text{if}~i<j\\
    0, & \text{otherwise}
    \end{cases}
\end{align}

Observe that the (usual $l_{\infty}$) operator norm of the matrix $\Delta_n$ is given explicitly by $ \| \Delta_n \|_{\infty}=\max_{1\leq i\leq n}H_{n,i}$,
where, for $1 \leq i\leq n-1$, $H_{n,i} ~\dot{=} (1+\Bar{\eta}_{i,i+1}+\cdots+\Bar{\eta}_{i,n})$.
For $i=n$, $H_{n,n}=1$  \cite{kontorovich2008concentration}. 

Using these definitions, the desired concentration result can be expressed as follows.

\begin{theorem}[Theorem 1.1 in \cite{kontorovich2008concentration}]\label{thm:app-a-1}
Suppose $\Sc$ is a countable space, $\Fc$ is the set of all subsets of $\Sc^n$, $\P$ is a probability measure on $(\Sc^n, \Fc )$ and $\varphi:\Sc^n\to \mathbb{R}$ is a $c$-Lipschitz function (with respect to the Hamming metric) on $\Sc^n$ for some $c > 0$. Then for any $t > 0$,
\begin{align}
\textstyle    \P\{\| \varphi - \E\varphi\|\geq t\}\leq 2\exp(-\frac{t^2}{2nc^2\|\Delta_n\|^2_\infty}).\label{eq:main-1-app1}
\end{align}

\end{theorem}

For the particular case when $(X_1, \ldots , X_n)$ is a (possibly inhomogeneous) Markov chain, the bound in Theorem \ref{thm:app-a-1} simplifies further. More precisely, given any initial probability distribution $p_0(\cdot)$ and stochastic transition kernels $p_i(\cdot|\cdot)$, $1\leq i\leq n-1$,  $1\leq i\leq n$, let the probability measure $\P$ on $\Sc^n$ be defined by
\begin{align}
\textstyle    \P\{(X_i, \ldots, X_i)=\xv\}=p_0(x_1)\prod^{i-1}_{j=1}p_j(x_{j+1}|x_j),\label{eq:Markov-def}
\end{align}
for any $1\leq i\leq n$ and any $\xv=(x_1,\ldots,x_i)\in\Sc^i$. Moreover, for $1\leq i\leq n-1$, let $\theta_i$ denote the $i$th contraction coefficient of the Markov chain defined as
\begin{align}
    \theta_i~~\dot{=} \sup_{x',x''\in\Sc}\| p_i(\cdot|x')-p_i(\cdot|x'')\|_{\rm TV}.
\end{align}
for $1\leq i\leq n-1$, and define
\begin{align}
    M_n~~\dot{=}\max_{1\leq i\leq n-1}(1+\theta_i+\theta_i\theta_{i+1}+\cdots+\theta_i\cdots\theta_{n-1}).\label{eq:def-Mn}
\end{align}
Then Theorem \ref{thm:app-a-1} can be simplified as follows. 
\begin{theorem}[Theorem 1.2 from \cite{kontorovich2008concentration}]\label{thm:app-a-2}
Suppose that $\P$ is the Markov measure on $\Sc^n$ described in \eqref{eq:Markov-def} and  $\varphi:\Sc^n\to \mathbb{R}$ is a $c$-Lipschitz function (with respect to the Hamming metric) on $\Sc^n$ for some $c > 0$. Then for any $t > 0$,
\begin{align}
\textstyle    &\P\{\| \varphi - \E\varphi \|\geq t\}
    \leq 2\exp\left(-\frac{t^2}{2nc^2M_n^2}\right),
\end{align}
where $M_n$ is defined in \eqref{eq:def-Mn}. 
\end{theorem}

\end{appendices}


\bibliographystyle{unsrt}
\bibliography{myrefs}

\end{document}

%% file: defns.tex
\usepackage{xspace}
\usepackage{bbm}
\newcommand{\Ac}{\mathcal{A}}
\newcommand{\Bc}{\mathcal{B}}
\newcommand{\Cc}{\mathcal{C}}

\newcommand{\Ec}{\mathcal{E}}
\newcommand{\Fc}{\mathcal{F}}

\newcommand{\Lc}{\mathcal{L}}

\newcommand{\Qc}{\mathcal{Q}}

\newcommand{\Sc}{\mathcal{S}}

\newcommand{\Av}{{\bf A}}
\newcommand{\Bv}{{\bf B}}
\newcommand{\Cv}{{\bf C}}
\newcommand{\Dv}{{\bf D}}
\newcommand{\Hv}{{\bf H}}
\newcommand{\Xv}{{\bf X}}
\newcommand{\Yv}{{\bf Y}}
\newcommand{\Zv}{{\bf Z}}

\newcommand{\xv}{{\bf x}}
\newcommand{\yv}{{\bf y}}
\newcommand{\zv}{{\bf z}}

\newcommand{\cv}{{\bf c}}
\newcommand{\dv}{{\bf d}}

\newcommand{\muv}{\boldsymbol \mu}

\DeclareMathOperator\E{E}
\let\P\relax
\DeclareMathOperator\P{P}

\def\textiid{i.i.d.\@\xspace}
\newcommand\iid{\ifmmode\text{ i.i.d. } \else \textiid \fi}

